\documentclass[12pt]{article}
\usepackage{amsmath,amsfonts,amssymb,array,verbatim,amscd}
\usepackage{graphicx,subfigure}
\usepackage{cancel}
\usepackage{color}
\usepackage{ifthen}

\makeatletter
\@addtoreset{equation}{section}

\renewcommand{\thefootnote}{\fnsymbol{footnote}}
\makeatother

\usepackage{fancyhdr}

\usepackage{pdflscape}
\newcommand{\NF}{N_{\rm f}}

\newcommand{\non}{\nonumber\\}
\newcommand{\D}{\mathcal{D}}
\newcommand{\p}{\partial}
\newcommand{\Tr}{{\rm Tr}}
\newcommand{\tr}{{\rm tr}}
\newcommand{\diag}{{\rm diag}}

\newcommand{\beq}{\begin{eqnarray}}
\newcommand{\eeq}{\end{eqnarray}}
\newcommand{\bpm}{\begin{pmatrix}}
\newcommand{\epm}{\end{pmatrix}}
\newcommand{\ba}{\left( \begin{array}}
\newcommand{\ea}{\end{array} \right)}
\newcommand{\be}{\begin{equation}}
\newcommand{\ee}{\end{equation}}
\newcommand{\bea}{\begin{eqnarray}}
\newcommand{\eea}{\end{eqnarray}}
\newcommand{\beann}{\begin{eqnarray*}}
\newcommand{\eeann}{\end{eqnarray*}}
\newcommand{\vs}[1]{\vspace{#1 mm}}
\newcommand{\hs}[1]{\hspace{#1 mm}}

\makeatletter
\@addtoreset{equation}{section}

\renewcommand{\thefootnote}{\fnsymbol{footnote}}
\makeatother

\def\de{\partial}
\def\Tr{ \hbox{\rm Tr}}

\def\diag{\hbox{\rm diag}}

\def\p{\partial}

\newcommand{\R}{\mathbb{R}}
\newcommand{\C}{\mathbb{C}}

\makeatletter
\def\mr@ignsp#1 {\ifx\:#1\@empty\else #1\expandafter\mr@ignsp\fi}%
\newcommand{\multiref}[1]{\begingroup
\xdef\mr@no@sparg{\expandafter\mr@ignsp#1 \: }%
\def\mr@comma{}%
\@for\mr@refs:=\mr@no@sparg\do{\mr@comma\def\mr@comma{,}\ref{\mr@refs}}%
\endgroup}
\makeatother

\begin{document}

\thispagestyle{empty}
\begin{flushright}
IFUP-TH/2012-13
\end{flushright}
\vspace{1mm}
\begin{center}%
{\LARGE \bf Effective Action of Non-Abelian \\ \vs{3}
Monopole-Vortex Complex} \\

\vspace{5mm}%
{\normalsize\bfseries Mattia Cipriani$^a$, ~~Toshiaki Fujimori$^{a,b}$}
\footnotetext{
Email addresses: \tt
mattia.cipriani(at)df.unipi.it, 
toshiaki.fujimori(at)pi.infn.it}

\vspace{5mm}
${}^a${\it\small
Department of Physics,``E. Fermi`'', University of Pisa,
Largo B.~Pontecorvo, 3, 56127 Pisa, Italy and}\\
{\it\small 
INFN, Sezione di Pisa,
Largo B.~Pontecorvo, 3, 56127 Pisa, Italy}
\vspace{2mm}

${}^b${\it\small
Department of Physics, National Taiwan University, Taipei 10617, Taiwan}

\par\vspace{1cm}

\vfill

\begin{abstract}
We construct effective actions for non-Abelian 1/4 Bogomol'nyi-Prasad-Sommerfield (BPS) monopole-vortex complexes in $4d$ $\mathcal{N}=2$ supersymmetric gauge theories with gauge groups $U(N)$, $U(1) \times SO(2n)$ and $U(1) \times USp(2n)$. In the color-flavor locked vacuum with degenerate hypermultiplet masses, a subgroup of the color-flavor diagonal symmetry remains unbroken and gives internal orientational moduli to vortices which confine monopoles in the Higgs phase. In this paper we discuss the effective action which describes the interactions between monopoles and the orientational moduli of non-Abelian vortices both from the bulk and vortex worldsheet theories. In the large mass limit, we find that the effective action consists of two-dimensional non-linear sigma models on vortex worldsheets and boundary terms which describes monopole-vortex interactions.
\end{abstract}

\vspace{3mm}

\vspace*{\fill}
\end{center}

\setcounter{footnote}{0}
\renewcommand{\thefootnote}{\arabic{footnote}}
\setcounter{page}{1}

\newpage

\section*{Introduction}

Magnetic monopoles have attracted many attentions in the context of non-perturbative QCD. They are believed to play a fundamental role in the confinement in the picture of dual superconductivity \cite{'tHooft:1981ht}. Non-Abelian monopoles, which exist when there is an unbroken non-Abelian gauge symmetry, would be particularly important if the dual magnetic system is non-Abelian \cite{Weinberg:1979zt}. However, there are well-known difficulties related to the transformation properties of non-Abelian monopoles under the unbroken gauge group \cite{Abouelsaood:1983gw}: one is the topological obstruction which prevents the existence of globally well-defined generators in the monopole background and the other is the non-normalizability of certain bosonic modes related to the gauge transformations. On the other hand, when monopoles are in the Higgs phase \cite{Tong:2003pz}, these modes become ``normalizable" in the sense that they are trapped inside vortices and become internal degrees of freedom living on the two-dimensional worldsheet of the non-Abelian vortices attached to the monopole.

Non-Abelian vortices were found in the color-flavor locked vacuum of ${\cal N}=2$ supersymmetric $U(N)$ gauge theory \cite{Hanany:2003hp} and there has been a lot of progress in the study of their properties \cite{review,Eto:2006pg}. Unlike Abelian vortices \cite{Abrikosov:1956sx}, they possess internal degrees of freedom, called {\itshape orientational moduli}, related to the breaking of the $SU(N)$ color-flavor diagonal symmetry by the vortex configuration itself. The effective dynamics of the orientational moduli of a single $U(N)$ non-Abelian vortex is described by the two-dimensional $\mathcal N =(2,2)$ $\C P^{N-1}$ sigma model \cite{Hanany:2003hp,Hanany:2004ea,Eto:2004rz,Gorsky:2004ad}. Besides the case of $U(N)$ gauge group, the analysis of non-Abelian vortices and their orientational moduli has also been extended to arbitrary gauge groups \cite{Eto:2008yi}. In particular, the cases of $U(1) \times SO(2n)$ and $U(1) \times USp(2n)$ has been extensively studied \cite{Ferretti:2007rp} and the effective worldsheet actions for the $SO(2n)/U(n)$ and $USp(2n)/U(n)$ orientational moduli were constructed in Ref.\,\cite{Gudnason:2010rm}. One of the most important facts about the orientational moduli of non-Abelian vortices is that the effective vortex worldsheet theory admits kink solutions corresponding to confined monopoles in the Higgs phase \cite{Tong:2003pz}. Monopoles appear as kinks in the vortex worldsheet theory: the system is in the Higgs phase and the magnetic flux spreading out from monopoles is carried away by the vortices connected to them. Such configuration of \textit{BPS monopole-vortex complex} provides a physical explanation \cite{Hanany:2004ea} for the relationship between BPS spectra in two-dimensional $\mathcal N=(2,2)$ sigma models and four-dimensional $\mathcal N = 2$ supersymmetric QCD \cite{Dorey:1998yh}. Recently the confined monopoles in the Higgs phase has been generalized to the case of $U(1) \times SO(2n)$ and $U(1) \times USp(2n)$ gauge theories \cite{Eto:2011cv}. 

The aim of this paper is to discuss the effective theory of the monopole-vortex complex which consists of non-Abelian vortices and monopoles. This is a continuation of the work done in a similar model \cite{Cipriani:2011xp}, a softly broken $\mathcal{N}=2$ $SU(N+1)$ theory, where a numerical calculation of the complex was carried out. In the sftly broken $\mathcal{N}=2$ model, however, the monopole-vortex complex is not a BPS state and it turned out that the construction of its effective model is difficult. In this paper, we study models with $\mathcal N =2$ supersymmetry where the monopole-vortex complex is a BPS configuration, to which the \textit{moduli matrix formalism} is applicable. The moduli matrix formalism for solitons in the Higgs phase \cite{Eto:2006pg} proved to be a very powerful tool in understanding the various aspects of solitons, such as their moduli spaces and low energy effective dynamics etc. By using this approach, effective theories of various solitons has been discussed: domain walls \cite{Isozumi:2004va}, non-Abelian vortices \cite{Eto:2004rz,Fujimori:2010fk}, domain wall networks \cite{Eto:2006bb} and vortex strings stretched between domain walls \cite{Eto:2008mf}. Recently non-Abelian monopoles in the Higgs phase have been classified by using the moduli matrix formalism \cite{Nitta:2010nd}. 

In this paper, we consider degenerate hypermultiplets masses which break the color-flavor diagonal symmetry $G_{C+F}$ to its non-Abelian subgroup $H_{C+F}$. In this case, the vortices attached to the monopoles can be non-Abelian vortices which have orientational zero modes arising from the unbroken symmetry $H_{C+F}$. By using the moduli matrix formalism, we discuss the effective action describing the interaction between monopoles and the orientational moduli of the non-Abelian vortices. If we assume that the mass scale of monopoles $m$ is much larger than the scale $\Lambda$ of the sigma model for the orientational moduli, we can obtain the action of non-linear sigma model interacting with the monopoles on the boundaries of the vortex worldsheets. Such an effective theory would be useful for the study of quantum physics of non-Abelian monopoles with zero modes arising due to non-Abelian symmetries. 

This paper is organized as follows. In Sec.\,\ref{TheModel}, we explain the features of our model and introduce the moduli matrix formalism for the BPS configurations. We begin in Sec.\,\ref{U2case} with the simplest configuration in the $U(2)$ case: two Abelian vortices connected by a monopole. Sec.\,\ref{U3case} is devoted to the first non-trivial example where non-Abelian vortices appear: $U(3)$ gauge group with $N_f=3$ flavors. In Sec.\,\ref{sec:eff} we explain our results for the effective actions of the monopole-vortex complexes. We discuss the effective actions by deriving them from the bulk theory in Sec.\,\ref{actionfrombulk} and from the vortex worldsheet theory in Sec.\,\ref{U3kinkaction}. In Sec.\,\ref{SOUSpcase} we discuss the case of $U(1) \times SO(2n)$ and $U(1) \times USp(2n)$ gauge groups by using the vortex worldsheet theory. Sec.\,\ref{Conclusions} is devoted to our conclusions and discussion for future work.

\section{The model}\label{TheModel}
First, we briefly review the BPS monopole-vortex complex 
in the ${\cal N}=2$ supersymmetric QCD with gauge group $U(1) \times G$ 
and $N_f$ matter hypermultiplets 
in the fundamental representation of the gauge group. 
In this paper, we consider the theories with $G=SU(N),\, SO(2n)$ and $USp(2n)$. For simplicity, we deal with only the case of $G=SU(N)$ in this section. 
The bosonic part of the supersymmetric action 
which is relevant to the monopole-vortex complex is
\begin{align}\label{theaction}
S= \int d^4x \, \Tr\left\{ \frac{1}{2g^2} F_{\mu\nu}^2 + \frac{1}{g^2} |\D_\mu \Phi|^2 + |{\cal D}_\mu Q|^2 - |\Phi Q - Q M |^2 - \frac{g^2}{4} (Q Q^\dagger - \xi)^2 \right\} \, ,
\end{align}
where $F_{\mu \nu} = \p_\mu A_\nu - \p_\nu A_\mu + i [ A_\mu , A_\nu]$ 
is the field strength, $\Phi$ is a real adjoint scalar and 
$Q$ is a $N$-by-$N$ matrix which represents $\NF=N$ scalar fields 
in the fundamental representation of the 
$U(N) \cong U(1) \times SU(N)$ gauge group. 
In this paper, we set the $N$-by-$N$ mass matrix to be real: 
$M^\dagger = M$. 
Using a flavor rotation, we can diagonalize the mass matrix as 
$M = \diag(m_1,\cdots,m_N )$. Note that the trace part of $M$ is unphysical since it can be absorbed by shifting $\Phi$.  

This model has a unique vacuum, 
defined by the following vacuum expectation values (VEVs) 
for the scalar fields:
\begin{equation}
	\langle \Phi \rangle = M \, , \ \ \ \ \ \ \ \ \langle Q \rangle = \sqrt{\xi} \, \mathbb{I}_N \, .
\end{equation}
This vacuum is invariant under a color-flavor locked global symmetry $H_{C+F}$ 
\begin{equation}
\Phi \stackrel{H_{C+F}}{\longrightarrow}H_C \, \Phi \, H_C^{-1}, \, \qquad 
Q \stackrel{H_{C+F}}{\longrightarrow} H_C \, Q \, H_F^{-1}, \, \qquad 
H_C=H_F \subset G \, .
\end{equation}
If all the masses are non-degenerate $m_i \not = m_j~(\mbox{for $i \not = j$})$, the unbroken symmetry is the Cartan subgroup 
$H_{C+F} = U(1)^{r}~(r=\text{rank} \, G)$. 
On the other hand, if some of the masses are equal, 
the global symmetry in the vacuum takes the form
\begin{equation}
H_{C+F} = S \left( U(n_1) \times U(n_2) \times \dots \times U(n_q) \right) \, ,
\end{equation}
with some integer $q$. 
Since there are two different mass scales in the VEVs, 
it is possible to consider two different 
hierarchical symmetry breaking patterns. 
In the case of $m \gg \sqrt{\xi}$, we have the following situation:
\begin{equation}
(U(1) \times G) \times SU(N)_F ~\stackrel{m}{\longrightarrow}~ (U(1) \times H_C) \times H_F ~\stackrel{\sqrt{\xi}}{\longrightarrow}~ H_{C+F} \, .
\end{equation}
The first symmetry breaking supports 
monopoles of size of order $m^{-1}$, 
confined by very wide flux tubes (vortex) 
of transverse width of order $\sqrt{\xi}^{-1}$. 
For reasons which will become clear later, 
we will consider the opposite case, namely $\sqrt{\xi} \gg m$. 
For this range of parameters, the symmetry breaking is the following:
\begin{equation}\label{SB}
(U(1) \times G) \times SU(N)_F ~\stackrel{\sqrt{\xi}}{\longrightarrow}~ G_{C+F} ~\stackrel{m}{\longrightarrow}~ H_{C+F}
\end{equation}
The flux tubes are very narrow, squeezing monopoles inside them. Monopoles now correspond to kinks interpolating between different vacua in the vortex worldsheet theory. In this paper, we focus on monopole-vortex complexes
constructed by non-Abelian vortices connected by monopoles. 
We would like to determine the effective action 
which describes the interaction between the monopoles 
and the orientational moduli of the non-Abelian vortices. 

In the supersymmetric theory, static monopole-vortex configurations 
are 1/4 BPS solutions which satisfy the first order BPS equations 
which can be obtained as follows: 
the energy of static configurations can be rewritten 
in the well known Bogomol'nyi form as
\begin{align}\label{BPSequations}
E = \int d^3 x \, \, \Tr &\left[ \phantom{+} \frac{1}{g^2} \left( F_{12} - {\cal D}_3\Phi + \frac{g^2}{2} \left( Q Q^\dagger - \xi \mathbf 1_N \right) \right)^2 \right. \non 
& \ \ + |\D_1 Q + i \D_2 Q|^2 + |\D_3 Q + \Phi Q - Q M|^2 \phantom{\bigg[} \non 
& \ \ + \frac{1}{g^2} \left( F_{23} - {\cal D}_1 \Phi \right)^2 + \frac{1}{g^2} \left( F_{13} - {\cal D}_2 \Phi \right)^2 \non
& \ \ \left. + \frac{1}{g^2} \de_i (\epsilon_{ijk} \Phi F_{jk}) + \xi F_{12} \right] \, ,
\end{align}
where we have ignored total derivative terms which do not contribute to the energy of the BPS monopole-vortex complex. The last two terms correspond to the topological charges of monopoles and vortices stretched along $x_3$-axis, respectively. Imposing the vanishing of the squared terms leads to the following BPS equations:
\begin{subequations}
	\begin{equation} \label{BPS1}
		\D_1 Q + i \D_2 Q ~=~ \D_3 Q + \Phi Q - Q M ~=~ 0 \, ,
	\end{equation}
	\begin{equation} \label{BPS2}
		F_{23} - {\cal D}_1 \Phi ~=~ F_{31} - {\cal D}_2 \Phi ~=~ 0 \, ,
	\end{equation}
	\begin{equation} \label{BPS4}
		F_{12} - {\cal D}_3\Phi + \frac{g^2}{2} \left( Q Q^\dagger - \xi \mathbf 1_N \right)^2 = 0 \, .
	\end{equation}
\end{subequations}
The general solution to Eqs.\,\eqref{BPS1} and \eqref{BPS2} can be written as
\begin{equation}
A_{\bar{z}} = i S^{-1} \de_{\bar{z}} S \, , \qquad 
A_3 - i \Phi = i S^{-1} \p_{3} S \, , \qquad
Q = \sqrt{\xi} \, S^{-1} H_0(z) e^{M x_3} \, ,
\label{eq:BPS_sol}
\end{equation}
where $z = x_1 +i \, x_2$. 
$H_0(z)$ is a $N$-by-$N$ holomorphic matrix called {\itshape moduli matrix}, 
while $S(z, \bar{z},x_3)$ is an element of the complexified gauge group $U(N)^\C \cong GL(N,\C)$, namely an invertible $N$-by-$N$ matrix. 
Then, the last BPS equation \eqref{BPS4} can be rewritten as \cite{Isozumi:2004vg}
\begin{equation}\label{mastereq}
\frac{1}{g^2 \xi} \Big[ 4 \de_z \left( \Omega \de_{\bar{z}} \Omega^{-1} \right) + \de_3 \left( \Omega \de_3 \Omega^{-1} \right) \Big] = \Omega_0 \Omega^{-1} - \xi \mathbf 1_N \, ,
\end{equation}
where we have defined hermitian matrices $\Omega$ and $\Omega_0$ by
\begin{equation} 
\Omega = S S^\dagger, \hs{10} \Omega_0= H_0 e^{2M \, x_3} H_0^\dagger \, .
\end{equation}
Eq.\,\eqref{mastereq} is called the master equation for 1/4 BPS configurations.
Since $QQ^\dagger \rightarrow \xi \mathbf 1_N$ at the spatial infinity, 
the boundary condition for $\Omega$ is given by
\beq
\Omega ~\underset{|z| \rightarrow \infty}{\longrightarrow}~ \Omega_0. \label{eq:boundary}
\eeq
Once we solve this equation with respect to $\Omega$ 
for a given moduli matrix $H_0$, 
we can determine the matrix $S$ 
up to the gauge transformation $S \rightarrow S U$ with $U \in U(N)$. 
Then, the BPS solution $(Q,\,A_\mu)$ can be obtained
through Eq.\,\eqref{eq:BPS_sol}. 
Therefore, the moduli matrix $H_0(z)$ classifies 
all the BPS configurations and 
the parameters contained in $H_0(z)$ can be 
identified with the moduli parameters of the BPS configurations.
Note that the solution Eq.\,\eqref{eq:BPS_sol} and the master equation 
\eqref{mastereq} are invariant under the following transformation 
called the $V$-transformation
\begin{equation}\label{Vequivalence}
S \longrightarrow V(z) S \, , \qquad H_0 \longrightarrow V(z) H_0 \, ,
\end{equation}
where $V(z)$ is an arbitrary holomorphic $N$-by-$N$ matrix 
which is an element of the complexified gauge group $U(N)^\C \cong GL(N,\C)$. 
Thus the matrices $S$ and $H_0$ are defined modulo this $V$-transformation, 
so that we have to fix this redundancy to find physical moduli parameters contained in the moduli matrix $H_0(z)$. 

For a given moduli matrix $H_0(z)$, the vortex number can be determined as follows. The vortex charge density can be rewritten in terms of $\Omega$ as
\beq
- \xi \, \tr F_{12} &=& \xi \p_z \p_{\bar z} \log \det \Omega, \phantom{\frac{1}{g^2}} \label{eq:vorticity}
\eeq
From this and the boundary condition \eqref{eq:boundary}, 
we can see that the number of vortices are determined by 
the degree of the polynomial $\det H_0$ 
\beq
\det H_0 ~=~ z^k + \cdots ~~~ \Longrightarrow~~~ - \xi \int d^2 x \, \tr \, F_{12} ~=~ 2 \pi \xi k.
\eeq
The zeros of the polynomial $\det H_0$ correspond to the positions of vortices.

In what follows, we consider configurations involving non-Abelian vortices and monopoles. The nature of the vortices depends crucially on the choice of the mass matrix. If all the masses are different, the residual global group is a product of $U(1)$ factors. In this case, all vortices are of the Abelian type since all the orientational zero modes are lifted by a potential induced by the non-degenerate masses. However, if some of $m_i$ are equal, the unbroken color-flavor group is a non-Abelian subgroup $H_{C+F}(\subset G_{C+F})$ and there exist non-Abelian vortices which have orientational zero modes arising due to the non-Abelian symmetry group $H_{C+F}$.

\subsection{Two Abelian vortices: $G=SU(2)$}\label{U2case}
Before discussing the monopole-vortex complex involving non-Abelian vortices, 
let us first review the simplest example in the case of $G=SU(2)$. 
In this case the mass matrix is given by
\beq
M = \frac{1}{2} \ba{cc} m & 0 \\ 0 & -m \ea.
\eeq
In the massless case $(m=0)$, a single 1/2 BPS vortex in this model has the orientational moduli $\C P^1$ which parameterize the continuous set of configurations with degenerate energy. On the other hand, the non-zero mass term gives the potential on the moduli space and only two discrete points on $\C P^1$ remains BPS-saturated. The corresponding moduli matrices are given by
\beq
H_0^+ = \ba{cc} z-z_0 & 0 \\ 0 & 1 \ea, \hs{10}
H_0^- = \ba{cc} 1 & 0 \\ 0 & z-z_0 \ea, 
\eeq
where the complex parameter $z_0$ denotes the vortex position.
The corresponding solutions of the master equation Eq.\,\eqref{mastereq} 
take the form
\beq
\Omega^+ = e^{M x_3} \ba{cc} e^{\psi} & 0 \\ 0 & 1 \ea, \hs{10}
\Omega^- = e^{M x_3} \ba{cc} 1 & 0 \\ 0 & e^{\psi} \ea,
\eeq
where $\psi$ is the profile function for a single Abelian vortex configuration,  which satisfies
\beq
4\p_z \p_{\bar z} \psi = g^2 \xi ( 1 - |z-z_0|^2 e^{-\psi} ), \hs{10} \psi ~\underset{|z| \rightarrow \infty}{\longrightarrow} ~ \log |z-z_0|^2. 
\eeq
Note that the physical fields Eq.\,\eqref{eq:BPS_sol} are independent of $x_3$ 
for the 1/2 BPS configurations. For example, the magnetic flux $B_3 = F_{12}$ takes the form
\beq
B_3^{\pm} = - \p_z \p_{\bar z} \psi \, ( \mathbf 1_2 \pm \sigma_3 ).
\label{eq:AV}
\eeq

The 1/4 BPS configuration of the monopole-vortex complex is a junction of these two vortex strings. Since their $\sigma_3$-components of the magnetic flux have opposite orientations, there exist a magnetic monopole at the junction point. 
We can construct 1/4 BPS configurations of monopole-vortex complex by using more general moduli matrix. Since $\det H_0$ is a polynomial of degree one for a single vortex ($k=1$), the generic moduli matrix $H_0$ can be fixed by using the $V$-transformation as
\beq\label{N2matrix}
H_0 = \ba{cc} z-z_0 & 0 \\ a & 1 \ea, \hs{10} z_0,\,b \in \C,
\eeq
where $z_0$ is the position of the vortex string in the complex $z$-plane. 
For later convenience, let us redefine the complex parameter $a$ 
as
\beq
a ~=~ \exp ( - m x_0 - i \eta ), \hs{10} x_0 \,, \eta \in \R. 
\eeq
In order to solve the master equation Eq.\,\eqref{mastereq} , it is convenient to rewrite $\Omega$ in terms of real functions $\psi_1, \psi_2$ and a complex function $\tau$ as
\beq
\Omega = e^{M (x_0 + i \eta / m)} \ba{cc} 1 & \tau \\ 0 & 1 \ea \ba{cc} e^{\psi_1} & 0 \\ 0 & e^{\psi_2} \ea \ba{cc} 1 & 0 \\ \bar \tau & 1 \ea e^{M (x_0 - i \eta / m )}. \label{eq:SU(2)_sol}
\eeq
We can expand $\psi_1, \psi_2$ and $\tau$ 
in terms of the ratio of the mass scales $m/(g \sqrt{\xi})$. 
By expanding the master equation Eq.\,\eqref{mastereq}, 
we find that the lowest order solutions are given by
\beq
\psi_1 &=& \psi - \log \big( e^{m (x_3-x_0)} + e^{-m (x_3-x_0)} \big) + \mathcal O \left( \frac{m^2}{g^2 \xi} \right), \label{eq:approx1} \\
\psi_2 &=& \hs{8} \log \big( e^{m (x_3-x_0)} + e^{-m (x_3-x_0)} \big) + \mathcal O \left( \frac{m^2}{g^2 \xi} \right), \label{eq:approx2}\\
\tau \ &=& (z-z_0) \frac{e^{m (x_3-x_0)}}{e^{m (x_3-x_0)} + e^{-m (x_3-x_0)}} + \mathcal O \left( \frac{m^2}{g^2 \xi} \right). \label{eq:approx3}
\eeq
From the matrix $\Omega (\equiv SS^\dagger)$, the matrix $S$ can be determined up to the $U(2)$ gauge transformation and then we can obtain the original fields $Q$ and $A_\mu$ (see Eq.\,\eqref{eq:BPS_sol}). In the singular gauge in which $Q \rightarrow \sqrt{\xi} \mathbf 1_2$ $(|z|\rightarrow \infty)$, the leading order solution takes the form   
\beq
Q &\approx& \sqrt{\xi} \, U^\dagger (x_3) \ba{cc} |z-z_0| e^{-\frac{1}{2} \psi} & 0 \\ 0 & 1 \ea U(x_3) , \\
A_{\bar z} &\approx& - \frac{i}{4} \left( \p_{\bar z} \psi - 1/\bar z \right) \, \Big[ \mathbf 1_2 + U^\dagger(x_3) \sigma_3 U(x_3) \Big] , \\
A_3 + i \Phi &\approx& -i M + \frac{i m \bar \beta}{1+ |\beta|^2} ( 1 - |z-z_0| e^{-\frac{1}{2} \psi} ) \, U^\dagger(x_3) \Big[ \sigma_1 + i \sigma_2 \Big] U(x_3),
\eeq
where $U(x_3) \in U(2)$ and the function $\beta(x_3)$ are defined by
\beq
U(x_3) = \frac{1}{\sqrt{1+|\beta|^2}} \ba{cc} 1 & - \bar \beta \\ \beta & 1 \ea, \hs{10} \beta(x_3) = \exp \left[ m ( x_3 - x_0 ) - i \eta \right]. 
\eeq
The magnetic flux is given by
\beq
B_3 ~\approx~ \p_z \p_{\bar z} \psi \, \Big[ \mathbf 1_2 + U^\dagger(x_3) \sigma_3 U(x_3) \Big].
\eeq
From this expression, we can see that this confiugration approaches to the 1/2 BPS vortex configurations in \eqref{eq:AV} at the spatial infinities $x_3 \rightarrow \pm \infty$. Since these two vortices are connected at $x_3 = x_0$, the real parameter $x_0$ can be identified with the monopole position. On the other hand, the parameter $\eta$ is the phase modulus of the monopole related to the $U(1)$ symmetry of the theory.

\subsection{Abelian and non-Abelian vortices: $G=SU(3)$}\label{U3case}
Next, we consider the simplest monopole-vortex complex involving a non-Abelian vortex in the case of $G=SU(3)$. Here, we take the following mass matrix
\beq
M = \frac{1}{2}
{\renewcommand{\arraystretch}{0.8}
{\arraycolsep .6mm
\ba{c|cc} m & & \\ \hline & -m & \\ & & -m \ea 
}}\, . \label{eq:massmatrix}
\eeq
As a consequence of this choice for the mass matrix, the residual global color-flavor group is $H_{C+F}=SU(2) \times U(1)$. In this case, there are two types of 1/2 BPS vortex configurations; one is Abelian and the other is non-Abelian. 
They are described by the following moduli matrices
\beq
{\renewcommand{\arraystretch}{0.8}
{\arraycolsep 1.5mm
H_0^A(z) = \ba{c|cc} z & & \\ \hline & 1 & \\ & & 1 \ea, \hs{10}
H_0^{NA}(z) = \ba{c|cc} 1 & & \\ \hline & z & \\ & b & 1 \ea}},
\eeq
where we have fixed the vortex position at $z=0$. 
Since the first moduli matrix does not have any parameter, 
it corresponds to the Abelian vortex configuration. 
On the other hand, the second moduli matrix, corresponding to the non-Abelian vortex, is parametrized by a complex parameter $b$. This matrix can be obtained from the moduli matrix with $b=0$ by using the $SU(2)$ symmetry as 
\beq
{\renewcommand{\arraystretch}{0.8}
{\arraycolsep 1.5mm
H_0^{NA}(z) ~=~ V(z) \ba{c|cc} 1 & & \\ \hline & z & \\ & & 1 \ea \ba{c|c} 1 & \\ \hline & \phantom{\bigg[} U \ea}},
\eeq
where the elements of the $V$-transformations and $SU(2)$ are respectively given by 
\beq
V(z) = \ba{c|cc} 1 & & \\ \hline & \frac{1}{\sqrt{1+|b|^2}} & \frac{\bar b z}{\sqrt{1+|b|^2}} \\ & 0 & \sqrt{1+|b|^2}\ea, \hs{10}
U = \frac{1}{\sqrt{1+|b|^2}} \ba{cc} 1 & - \bar b \\ b & 1 \ea.
\eeq
The form of the matrix $U$ implies that the parameter $b$ is a moduli parameter associated with the $SU(2)$ symmetry broken to $U(1)$ subgroup by the vortex configuration. Thus, the parameter $b$ can be interpreted as the inhomogeneous coordinate of $\C P^1 \cong SU(2)/U(1)$. 

\begin{figure}[htbp]
\begin{center}
\includegraphics[width=170mm]{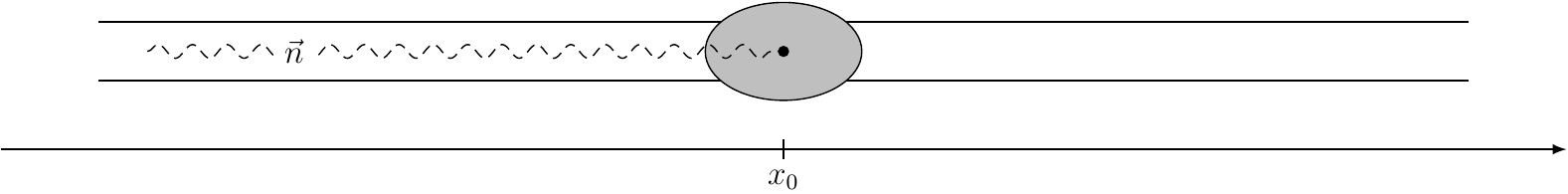}
\caption{\footnotesize The junction between a monopole and two different vortices: an Abelian vortex on the right side and a non-Abelian vortex on the left side. Orientational moduli traveling along the non-Abelian vortex end at the monopole.}
\label{ANApicture}
\end{center}
\end{figure}

It is possible to construct a junction of the Abelian vortex and the non-Abelian vortex with $b=0$ by embedding the solution for the $SU(2)$ monopole-vortex complex discussed in the previous subsection. The corresponding moduli matrix is given by
\beq
H_0 =
\ba{ccc}
z & 0 & 0 \\
a & 1 & 0 \\
0 & 0 & 1
\ea\, ,
\label{eq:b=0}
\eeq
where $a = \exp( - m x_0 - i \eta)$ and $x_0$ and $\eta$ are the monopole position and phase. By applying the $SU(2)$ transformation to Eq.\,\eqref{eq:b=0}, we can obtain the moduli matrix for the complex composed of the Abelian vortex and the non-Abelian vortex with $b \not = 0$ as
\beq
H_0(z) ~=~ 
\ba{c|c} 1 & \\ \hline & \phantom{\bigg[} U^\dagger \ea 
\ba{ccc}
z & 0 & 0 \\
a & 1 & 0 \\
0 & 0 & 1
\ea
\ba{c|c} 1 & \\ \hline & \phantom{\bigg[} U \ea 
~=~
\ba{ccc}
z & 0 & 0 \\
b_1 & 1 & 0 \\
b_2 & 0 & 1
\ea \, , 
\label{modulimatrix}
\eeq
where we have defined the parameters $b_1$ and $b_2$ by
\beq
b_1 = \frac{a}{\sqrt{1+|b|^2}}, \hs{10} 
b_2 = \frac{-ab}{\sqrt{1+|b|^2}}.
\eeq
The moduli matrix Eq.\,\eqref{modulimatrix} shows that 
the moduli space of the 1/4 BPS configuration\footnote{
The point $b_1=b_2=0$ is subtracted since it corresponds to the 1/2 BPS configuration of the Abelian vortex.} is $\mathcal M \cong \C^2 \setminus \{(0,0)\} \cong \R \times S^3$, where $\R$ is the monopole position and $S^3$ is the Hopf fibration of the phase and the orientational moduli $\C P^1$. 
Since the moduli matrix Eq.\,\eqref{modulimatrix} is essentially equivalent to that for the $SU(2)$ case, the corresponding solution $\Omega$ can be obtained by embedding the $SU(2)$ solution Eq.\,\eqref{eq:SU(2)_sol} as 
\beq
\Omega = V \ba{c|c} \Omega_{SU(2)} \phantom{\bigg[} \hs{-3} & \\ \hline & e^{-m x_3} \ea V^\dagger, \hs{10}
V = \ba{c|c} 1 & \\ \hline & \phantom{\bigg[} U^\dagger \ea .
\label{eq:Omega}
\eeq
As shown in Fig.\,\ref{ANApicture}, this configuration consists of Abelian and non-Abelian vortices connected by a monopole at $x_3=x_0$. Since the Abelian vortex is invariant under the $SU(2)$ symmetry, it does not have internal degrees of freedom and the orientational zero modes are localized only on the worldsheet of the non-Abelian vortex. 

As in the cases of other solitons, we can discuss the effective dynamics of the monopole-vortex complex by assuming that the moduli parameters are dynamical valuables depending on the time $t$. In general, effective dynamics is described by a non-linear sigma model whose target space is the soliton moduli space endowed with a metric. However, we cannot define a metric for the orientational moduli in this case since the non-Abelian vortex has semi-infinite worldvolume and hence the orientational zero modes are non-normalizable. The fact that the orientational zero modes are non-normalizable implies that there exist a continuous spectrum of the fluctuation modes parametrized by the momentum in $x_3$-direction. Therefore, the zero modes cannot be separated from the continuous massive modes and we have to take into account fluctuations of the orientational moduli propagating along the $x_3$-direction. In the next section, we discuss the effective action for the monopole-vortex complex by promoting the moduli parameters to fields depending on both $t$ and $x_3$.

\section{The effective action}\label{sec:eff}
\subsection{From the bulk theory}\label{actionfrombulk}

In this section, we discuss the effective action which describes the interactions between monopoles and the orientational moduli of non-Abelian vortices. In order to obtain a general form of the effective action, we do not deal with a specific moduli matrix and simply assume that $H_0(z)$ is a holomorphic function of the moduli parameters $\phi^i$, that is, the complex coordinates of the moduli space. Since the orientational moduli can propagate along the non-Abelian vortex, we assume that the moduli parameters $\phi^i$ has the $x_3$-dependence in addition to the time dependence. In other words, the moduli parameters contained in the BPS configurations are promoted to fields on the vortex worldsheet
\beq
\phi^i ~\longrightarrow~ \phi^i(t,x_3). 
\eeq
The $(t,x_3)$-dependence of the moduli parameters $\phi^i$ induces small deviations from the background BPS configuration. They can be determined from the equations of motion by using the derivative expansion with respect to the derivative $\p_\alpha~(\alpha = t, x_3)$. The lowest order correction is $\mathcal O(\p_\alpha)$ and can be determined from the Gauss' law equations
\begin{equation}\label{gauss}
\frac{2}{g^2} {\cal D}_i F_{i \alpha} = \frac{i}{g^2} \Big\{ \left[\Phi, ({\cal D}_\alpha \Phi)^\dagger \right] - \left[({\cal D}_\alpha \Phi), \Phi^\dagger \right] \Big\} - i \Big\{ ({\cal D}_\alpha Q) Q^\dagger - Q ({\cal D}_\alpha Q)^\dagger \Big\} \, ,
\end{equation}
where $\alpha=0,3$ and $i = 1, 2, 3$. The solution for $A_0$ is given by
\beq
A_{0} &= i \left( \delta_0 S^\dagger S^{\dagger \, -1} - S^{-1} \bar{\delta}_{0} S \right) \, ,
\eeq
where we have introduced the derivative operators $\delta_0$ and $\bar{\delta}_0$ by
\beq
\delta_0 = \frac{\p \phi^i}{\p t} \frac{\p}{\p \phi^i}, \hs{10} 
\bar{\delta_0} = \frac{\p \bar{\phi}^i}{\p t} \frac{\p}{\p \bar{\phi}^i}.
\eeq
On the other hand, it is difficult to solve the equation of motion for $A_3$ because of the explicit $x_3$-dependence of the background BPS configuration. However, if $g \sqrt{\xi} \gg m$, we can obtain the following approximate solution for $A_3$ 
\begin{align}
A_{3} &= \frac{i}{2} \left( \de_{\dot{3}} S^\dagger S^{\dagger \, -1} - S^{-1} \de_{\dot{3}} S \right) + i \left( \delta_3 S^\dagger S^{\dagger \, -1} - S^{-1} \delta_{\bar{3}} S \right) + O\left( \frac{m^2}{g^2 \xi} \right) \, ,
\end{align}
where the differential operator $\de_{\dot{3}}$ acts only on the explicit $x_3$-dependence and $(\delta_3, \bar{\delta}_{3})$ are defined by
\begin{align}
\delta_3 = \frac{\de \phi^i}{\de x_3}\frac{\de}{\de \phi^i} \, , \quad 
\bar{\delta}_{3} = \frac{\de \bar{\phi}^i}{\de x_3}\frac{\de}{\de \bar{\phi}^i} \, .
\end{align}
In the following, we will always assume that $g \sqrt{\xi} \gg m$ and use the lowest order approximation with respect to $m^2/(g^2 \xi)$. At the lowest order in $\p_\alpha$ and $m^2/(g^2 \xi)$, there is no modification to $A_z$, $\Phi$ and $Q$. Having determined the modified fields, we can now write down the effective action by substituting the modified fields into the original action. The terms which do not contain $\p_\alpha \phi^i$ and $\p_\alpha \bar{\phi}^i$ correspond to the energy density of the background BPS configuration. On the other hand, the terms which are quadratic in $\p_\alpha \phi^i$ and $\p_\alpha \bar{\phi}^i$ give the effective action and can be cast in the following form
\beq
S_{eff} = \xi \int d^4 x \, \bar{\delta}^\alpha \tr \left[ \delta_\alpha \Omega_0 \Omega^{-1} \right] + \mathcal O \left( \frac{m^2}{g^2 \xi} \right).
\label{eq:formula}
\eeq
This is the general formula for the effective action which describes the dynamics of the monopole-vortex complex. Once we specify the moduli matrix $H_0$ and the solution of the master equation $\Omega$, we can obtain the effective action by substituting them into Eq.\,\eqref{eq:formula} and integrating over $x_1$ and $x_2$ directions ($z$-plane). 

Now let us apply the general formula Eq.\,\eqref{eq:formula} for the $SU(3)$ monopole-vortex complex specified by Eqs.\,\eqref{modulimatrix} and \eqref{eq:Omega}. Since we are dealing with the lowest order terms in $m^2/(g^2 \xi)$, we can use the approximate solution Eqs.\,\eqref{eq:approx1}-\eqref{eq:approx3} for the $SU(2)$ solution $\Omega_{SU(2)}$ contained in $\Omega$. Then, the general formula Eq.\,\eqref{eq:formula} gives
\begin{equation}\label{beffaction}
S_{eff} = C \int dx_3 dt \frac{(e^{-2m \, x_3} + |\vec{b}|^2)\left( \de_\alpha \vec{b}^\dagger \cdot \de_\alpha \vec{b} \right) - | \vec{b}^\dagger \cdot \de_\alpha \vec{b} |^2}{\left( e^{-2m \, x_3} + | \vec{b} |^2\right)^2} + \mathcal O \left( \frac{m^2}{g^2 \xi} \right) \, ,
\end{equation}
where $\vec b = (b_1,b_2)$ and the constant $C$ is given by
\begin{equation}\label{Cconstant}
\displaystyle C ~=~ \xi \int dz d\bar{z} \, (1 - |z|^2 e^{-\psi} ) ~=~ \frac{4 \pi}{g^2} \, .
\end{equation}
To extract the physical meaning of the effective action, let us make the following reparametrization of the moduli parameters
\begin{align}\label{repar}
\vec b = e^{- m x_0} \, \vec n, \hs{10} 
\vec n = \frac{e^{-i \eta}}{\sqrt{1+|b|^2}} \ba{c} 1 \\ - b \ea,
\end{align}
where the unit vector $\vec n$ represents $S^3$ parametrized by the phase $\eta$ and the orientational moduli $b$. After this reparametrization, we obtain the following form of the effective Lagrangian
\begin{align}
\mathcal L_{eff} = \frac{\pi}{g^2} \Bigg[ {\rm sech}^2 \left[ m(x-x_0) \right] \Big[ m^2 \dot x_0^2 + |\vec n^\dagger \cdot \p_\alpha \vec n|^2 \Big] + 2 e^{m(x-x_0)} {\rm sech} \left[ m(x-x_0) \right] \, |D_\alpha \vec n|^2 \Bigg] \, ,
\label{eq:seff}
\end{align}
where the differential operator $D_\alpha$ is defined by
\beq\label{eq:covderivative}
D_\alpha = \p_\alpha - \vec n^\dagger \cdot \p_\alpha \vec n \, .
\eeq 
If we assume that the wave length of the fluctuation of the moduli parameters is much longer than $m^{-1}$ and $(g \sqrt{\xi})^{-1}$, we can take the limit $m,\, g \sqrt{\xi} \rightarrow \infty$, in which the monopole becomes a point-like object. The asymptotic form of the functions in Eq.\,\eqref{eq:seff} in the limit $m \rightarrow \infty$ are given by
\beq
{\rm sech}^2 \left[ m(x-x_0) \right] \rightarrow \frac{2}{m} \delta(x-x_0), \hs{10} 2 e^{m(x-x_0)} {\rm sech} \left[ m(x-x_0) \right] \rightarrow \theta(x-x_0).
\eeq
By taking the large mass limit, we obtain the final form for the effective action
\beq\label{naction}
S_{eff} &=& \frac{4\pi}{g^2} \int dt dx \, \Big[ \mathcal L_M + \mathcal L_V \Big] + \mathcal O \left( \frac{m^2}{g^2 \xi} \right) \, ,
\eeq
with
\beq
\mathcal L_M &=& \left( \frac{m}{2} \dot x_0^2 + \frac{1}{2m} |\vec n^\dagger \cdot \p_\alpha \vec n|^2 \right) \delta(x-x_0) \, , \\
\mathcal L_V \, &=& |D_\alpha \vec n|^2 \ \theta(x-x_0) \, . \vphantom{\bigg[}
\eeq
As we can see, this action is made of two parts: $\mathcal L_V$ is the standard action for the $\mathbb{C}P^1$ sigma model on the worldsheet of the non-Abelian vortex and $\mathcal L_M$ corresponds to the kinetic term for the monopole modulus and the interaction term between the monopole and the orientational moduli of the non-Abelian vortex.

To see the physical meaning of the effective action, let us consider the following classical equations of motion derived from Eq.\,\eqref{naction} 
\beq
0 &=& \p_\alpha \Big[ (\vec n^\dagger \cdot  \p^\alpha \vec n) \, \delta(x-x_0) \Big] \, , \\
m \ddot x_0 &=& \left[ - |D_\alpha \vec n|^2 - \frac{1}{m} (\vec n^\dagger \cdot \p_\alpha \vec n) \p_x (\vec n^\dagger \cdot \p^\alpha \vec n) \right]_{x=x_0} \, , \label{eq:position} \\
0 &=& \Big[ D_\alpha^2 + |D_\alpha \vec n|^2 \Big] \, \vec n \ \theta(x-x_0) + \Big[ \frac{1}{m} ( \vec n^\dagger \cdot \p_\alpha \vec n ) + \p_\alpha (x-x_0) \Big] D^\alpha \vec n \ \delta(x-x_0) \, . \label{eq:orientation}
\eeq
The first equation corresponds to the conservation law for the overall $U(1)$ symmetry and implies that
\beq
\Big[ \vec n^\dagger \cdot \p_t \vec n \Big]_{x=x_0} =~ imQ \, , \hs{10}
\Big[ \vec n^\dagger \cdot \p_x \vec n \Big]_{x=x_0} =~-imQ \, \dot x_0 \, ,
\label{eq:phase}
\eeq
where the constant $Q$ is the conserved charge of the overall $U(1)$ symmetry. 
Note that the effective action is invariant under the $U(2)$ symmetry which acts on the unit vector $\vec n$ as
\beq
\vec n \rightarrow U \vec n, \hs{10} U \in U(2) \, .
\eeq
The corresponding current takes the form 
$J_\alpha = J_{\alpha}^M + J_{\alpha}^V$, 
where $J_{\alpha}^M$ and $J_{\alpha}^V$ are $2$-by-$2$ matrices given by
\beq
J_{\alpha}^V &=& - \frac{i}{2} \Big[ D_\alpha \vec n \, \vec n^\dagger - \vec n \, (D_\alpha \vec n)^\dagger \Big] \theta(x-x_0) \, , \\
J_{\alpha}^M &=&  - \frac{i}{m} (\vec n^\dagger \cdot \p_\alpha \vec n) \, \vec n \, \vec n^\dagger \delta(x-x_0) \, .
\eeq
The vortex part $J_\alpha^V$ is the $SU(2)$ current for the $\C P^1$ sigma model and does not contain the overall $U(1)$ part. On the other hand, the monopole part $J^M_{\alpha}$ can be rewritten by using Eq.\,\eqref{eq:phase} as
\beq
J^{M}_t = Q \, \vec n \, \vec n^\dagger \, \delta(x-x_0), \hs{10} J_{x}^M = - \dot x_0 Q \, \vec n \, \vec n^\dagger \, \delta(x-x_0) \, ,
\eeq
The conserved charge $Q$ is the conjugate momentum for the monopole phase. 
In the semi-classical treatment of the monopole moduli, the conserved charge $Q$ takes integer values, corresponding to the infinite tower of dyons. The other equations of motion Eqs.\,\eqref{eq:position}, \eqref{eq:orientation} imply that the interaction between the orientational moduli and the monopole becomes non-trivial when the charge $Q$ is non-zero. 

\subsection{From the vortex worldsheet theory}\label{U3kinkaction}

In this section, we derive the effective action of the monopole-vortex complex by a different viewpoint, which will allows us to obtain the form of the action in a more direct way. It is known that the effective action for the orientational moduli of a single $U(N)$ vortex is described by the $\C P^{N-1}$ sigma model. The degenerate bulk mass term breaks the $SU(N)_{C+F}$ symmetry of the $\C P^{N-1}$ sigma model to the subgroup $H_{C+F}$ and induces a potential on the vortex moduli space $\C P^{N-1}$. In the case of $m \ll g^2 \xi$, the potential is given by the squared norm of the Killing vector corresponding to the isometry generated by $M \in SU(N)_{C+F}$. This massive $\C P^{N-1}$ sigma model has several disjoint components of vacua corresponding to the fixed-point set of the isometry.  Each vacuum component is not necessarily an isolated point: isolated points are Abelian vortices and continuous vacua correspond to non-Abelian vortices. In the case of $m \ll g^2 \xi$, monopoles connecting those vortices are well-approximated by 1/2 BPS kinks interpolating the disjoint components of vacua. Here we show that the same effective action Eq.\,\eqref{naction} can be obtained from the kink solution in the vortex effective theory. 

We start from the effective action of a single $SU(3)$ vortex, i.e. the massive $\mathbb{C}P^2$ sigma model. In terms of a three-component unit vector $\vec \phi$, the vortex effective action is expressed as
\begin{equation}\label{cp2action}
S ~=~ \frac{4\pi}{g^2} \int \text{d}^2x \Big[ \left( |\p_\alpha \vec \phi|^2 - |\vec \phi^\dagger \cdot \p_\alpha \vec \phi|^2 \right) - \left(|M \vec \phi|^2 - |\vec \phi^\dagger \cdot M \vec \phi|^2 \right) \Big]  \, ,
\end{equation}
where $M$ is the mass matrix given in Eq.\,\eqref{eq:massmatrix}.
The Abelian vortex correspond to $\vec \phi = (1,0,0)$ and the non-Abelian vortex with the orientational moduli $b$ is $\vec \phi \propto ( 0 , 1 , -b )$. The 1/2 BPS kink solution  is given by
\beq
\vec \phi = \frac{1}{\sqrt{e^{m(x_3-x_0)} + e^{-m(x_3-x_0)}}} e^{M (x_3-x_0)} \ba{c} 1 \\ \vec n \ea, \hs{10} \vec n^\dagger \cdot \vec n = 1,
\label{eq:kink_sol}
\eeq
where $x_0$ is the monopole position and the two-component unit vector $\vec n$ parametrize $S^3$ corresponding to the phase $\eta$ and the orientational moduli $b$. Now let us promote the moduli parameters to dynamical variables
\beq
x_0 \rightarrow x_0(t), \hs{10} \vec n \rightarrow \vec n(t,x_3). 
\label{eq:replacement}
\eeq
In this case, we can show that the deviations of the background fields 
induced by the dynamical moduli parameters are higher order corrections in the derivative expansion with respect to $\p_\alpha$. Therefore, we can obtain the lowest order effective action just by substituting the solution Eq.\,\eqref{eq:kink_sol} with the replacement Eq.\,\eqref{eq:replacement} into the $\mathbb{C}P^2$ action Eq.\,\eqref{cp2action}. Consequently, we obtain the effective action which precisely coincides with Eq.\,\eqref{eq:seff}. 

\begin{figure}[tbp]
\begin{center}
\includegraphics[width=1\linewidth]{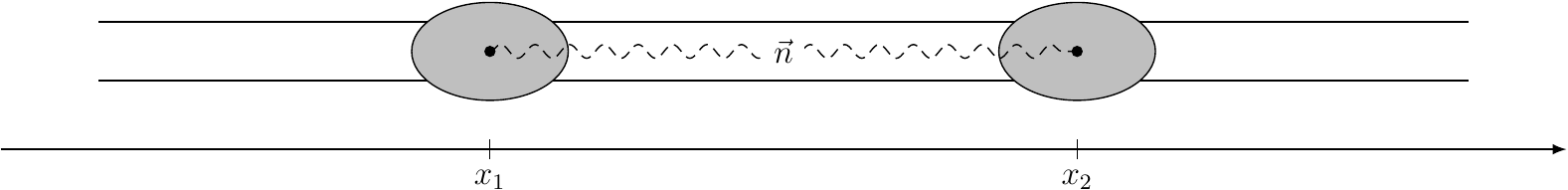}
\caption{\footnotesize A non-Abelian vortex stretched between two monopoles, connecting it to other two Abelian vortices. The corresponding effective action is the one of Eq. \eqref{eq:effactionmeson}.}
\label{NANApicture}
\end{center}
\end{figure}

Using the kink solution proves to be effective when we deal with more complicated setups. As an example, let us consider the $U(4)$ case with the following mass matrix which breaks the $SU(4)_{C+F}$ symmetry  to $SU(2) \times U(1)$:
\begin{equation}
M= \frac{1}{2}
{\renewcommand{\arraystretch}{0.8}
{\arraycolsep 1.3mm
\begin{pmatrix}
m & & & \\
  & 0 & & \\
  & & 0 & \\
  & & & -m
\end{pmatrix}}} \, . \label{eq:mass4}
\end{equation}
The $\C P^3$ effective action for the vortex orientation (four component vector $\vec \phi$) takes the same form as Eq.\,\eqref{cp2action}. In this case, there exist three types of BPS vortices: one is a non-Abelian vortex which has the $\C P^1$ orientational moduli parameter and the others are Abelian vortices. Therefore, we can construct a monopole-vortex complex which consists of the three vortices connected by two monopoles (see Fig.\,\ref{NANApicture}). 
The corresponding kink solution for the four component unit vector is given by\footnote{
In the following, we denote the spatial worldsheet coordinate by $x$ instead of $x_3$ for notational simplicity. } 
\begin{equation}\label{eq:kink_solU4}
\vec{\phi} = \frac{1}{\sqrt{e^{m (x-x_1)}+1+e^{-m (x-x_2)}}} e^{M x} \ba{cc} e^{-\frac{m}{2} x_1 + i \eta} \\ \vec n \\ e^{\frac{m}{2} x_2 - i \eta} \ea \, , \hs{10} \vec n^\dagger \cdot \vec n = 1,
\end{equation}
where $x_1$ and $x_2$ are the monopole positions and $\eta$ is the overall monopole phase, which is related to the $U(1)$ symmetry generated by the mass matrix \eqref{eq:mass4}. The two-component unit vector $\vec n$ parametrizes $S^3$ whose $U(1)$ fiber and $\C P^1$ base are the relative phase and the orientational moduli, respectively. 
Then we obtain
\beq
S_{eff} &=& \frac{4\pi}{g^2} \int dx dt \, \bigg[ f_1(x) \left( \frac{m^2}{4} \dot x_1^2 + |\vec n_1^\dagger \cdot \p_\alpha \vec n_1|^2 \right) +  f_2(x) \left( \frac{m^2}{4} \dot x_2^2 + |\vec n_2^\dagger \cdot \p_\alpha \vec n_2|^2 \right) \notag \\
&& \hs{18} + f_3(x) | D_\alpha \vec n |^2 - f_4(x) \left( \frac{m^2}{4} (\dot x_1 - \dot x_2)^2 + 4 |\vec n^\dagger \cdot \p_\alpha \vec n|^2 \right) \bigg]
\label{eq:U4_0}
\eeq
where we have defined $\vec n_1 = e^{-i \eta} \vec n$ and $\vec n_2 = e^{i \eta} \vec n$. The functions $f_i(x)$ and their asymptotic form in the limit $m \rightarrow \infty$ are given by
\beq
f_1(x) &=& \frac{e^{m (x-x_1)}+2e^{-m (x_1-x_2)}}{(e^{m (x-x_1)}+1+e^{-m (x-x_2)})^2}~\rightarrow~\frac{1}{m} \delta(x-x_1), \\
f_2(x) &=& \frac{e^{-m (x-x_2)}+2e^{-m (x_1-x_2)}}{(e^{m (x-x_1)}+1+e^{-m (x-x_2)})^2}~\rightarrow~ \frac{1}{m} \delta(x-x_2), \\
f_3(x) &=& \frac{1}{e^{m (x-x_1)}+1+e^{-m (x-x_2)}} ~~~~\, \rightarrow~ \theta(x-x_2)\theta(x_1-x), \\
f_4(x) &=& \frac{e^{-m (x_1-x_2)}}{(e^{m (x-x_1)}+1+e^{-m (x-x_2)})^2}~\rightarrow~ \mathcal O(e^{-m(x_1-x_2)}).
\eeq
Therefore, in the large mass limit $m \rightarrow \infty$, the effective action reduces to 
\beq\label{eq:effactionmeson}
S_{eff} &=& \frac{4\pi}{g^2} \int dx dt \, \bigg[ \frac{1}{m} \delta(x-x_1) \left( \frac{m^2}{4} \dot x_1^2 + |\vec n_1^\dagger \cdot \p_\alpha \vec n_1|^2 \right) + \frac{1}{m} \delta(x-x_2) \left( \frac{m^2}{4} \dot x_2^2 + |\vec n_2^\dagger \cdot \p_\alpha \vec n_2|^2 \right) \notag \\
&& \hs{18} + \theta(x-x_2)\theta(x_1-x) | D_\alpha \vec n |^2 + \mathcal O(e^{-m(x_1-x_2)}) \bigg]. \label{eq:U4}
\eeq
As this example shows, we can easily obtain the effective action in the parameter region $m \ll g^2 \xi$. However, since the vortex effective action does not have information which is higher order in $m^2/(g^2 \xi)$, it is impossible to determine the higher order corrections. Therefore, we have to start from the bulk action when we would like to determine the higher order corrections to the effective action. 

The extension to the case of general $N$ is straightforward. We can also discuss the generalization to the theories with $U(1) \times G$ gauge group. In the next section, we discuss the effective action for the monopole-vortex complex in $SO/USp$ gauge theories. 

\section{Monopole-Vortex complex in $SO/USp$ gauge theories}\label{SOUSpcase}
So far, we have discussed the monopole-vortex complex in the case of $U(N)$ gauge group. In this section, we consider generalization to the case of $U(1) \times G~(G=SO(2n),USp(2n))$ gauge group with $\NF = 2n$ flavors. In the following, we use the basis in which the elements in $G$ satisfy
\beq
U^T J U = J, \hs{10} J \equiv \ba{c|c} & \mathbf 1_n \\ \hline \pm \mathbf 1_n & \ea, \hs{10} U \in G,
\eeq
where plus and minus signs are for $SO(2n)$ and $USp(2n)$, respectively. 
In the massless theory, there exists a color-flavor locked vacuum $\langle Q \rangle \propto \mathbf 1_{2n}$ in which the diagonal color-flavor $G$ symmetry is not broken. In the massive theory, the vacuum with the VEV $\langle Q \rangle \propto \mathbf 1_{2n}$ remains if the mass matrix is an element of the Lie algebra of $G$, that is, 
\beq
M = \ba{c|c} M_n & \\ \hline & -M_n \ea, 
\eeq
where $M_n$ is a $n$-by-$n$ diagonal matrix. Due to this mass matrix, the diagonal color-flavor symmetry $G$ is broken to its subgroup. Here, we consider the case where the mass matrix is given by
\beq
M_n = \frac{m}{2} \mathbf 1_N.
\eeq
In this case, the unbroken subgroup is $U(n)$ whose elements take the form
\beq
U = \ba{c|c} U_n & \\ \hline & U_n^\ast \ea,
\eeq 
where $U_n$ is an $n$-by-$n$ unitary matrix and $U_n^\ast$ denotes the complex conjugate. Let us discuss the effective action for the monopole-vortex complex in this setup. Here, we use the sigma model description for the vortex effective action as in the previous section. The orientational moduli space of a single non-Abelian (local) vortex is 
\beq
\mathcal M_{orientation} \cong \left\{
\begin{array}{ll} 
SO(2n)/U(n) & \mbox{for $G = SO(2n)$} \\
USp(2n)/U(n) & \mbox{for $G = USp(2n)$} 
\end{array}. \right.
\eeq
The non-linear sigma model describing the dynamics of 
the non-Abelian vortex in the mass deformed theory is discussed in \cite{Gudnason:2010rm}.
The Lagrangian takes the form
\beq
\mathcal L_{eff} &=& \frac{4\pi}{g^2} \tr \Big[ ( \mathbf 1_n + B^\dagger B)^{-1} \p_\alpha B^\dagger ( \mathbf 1_n + B B^\dagger )^{-1} \p^\alpha B \notag \\
&& \hs{10} - ( \mathbf 1_n + B^\dagger B)^{-1} \{M_n, B^\dagger \} ( \mathbf 1_n + B B^\dagger )^{-1} \{M_n, B \} \Big].
\label{eq:USpL}
\eeq
The BPS equation and the kink solution are given by 
\beq
\p_x B = \{ M_n , B \}~~\Longrightarrow~~ B = e^{M_n x} B_0 e^{M_n x},
\label{eq:USp_kink}
\eeq
where $B_0$ is a constant (anti-)symmetric matrix
corresponding to the moduli parameters of the kink configuration. 
They are related to the monopole positions (and phases), which can be read from the energy density for the BPS kink configuration
\beq
\mathcal E ~=~ \frac{2\pi}{g^2} \p_x^2 \log \det (1 + B^\dagger B) ~=~ \frac{2\pi}{g^2} \p_x^2 \log \det (e^{-2 M_n x} + B_0^\dagger e^{2 M_n x} B_0) .
\label{eq:Esosup}
\eeq

Now, let us concentrate on the case of $M_n = m \mathbf 1_n$. 
In this case, the number of the disjoint components of vacua is $[\frac{n}{2}]+1$ for $SO(2n)$ and $n+1$ for $USp(2n)$. They correspond to the non-Abelian vortices whose orientational moduli spaces take the form
\beq
\mathcal M_{orientation} \cong \frac{SU(n)}{SU(p) \times SU(n-p) \times U(1)}, 
\eeq
with some integer $p$. To extract the physical meaning of the moduli parameters, it is convenient to reparameterize the moduli $B_0$ in the following way;
\beq
B_0 = U X_G U^T, \hs{10} U \in U(n), 
\label{eq:X}
\eeq
where $U$ is a unitary matrix and $X_G$ is the matrix given by
\beq
X_{SO(2n)}= \ba{ccc} e^{-m x_1} i \sigma_2 & & \\ & \ddots & \\ & & e^{-m x_{[n/2]}} i \sigma_2 \ea, \hs{5}
X_{USp(2n)} = \ba{ccc} e^{-m x_1} & & \\ & \ddots & \\ & & e^{-m x_n} \ea. 
\eeq
Note that the last component in $X_{SO(2n)}$ is zero if $n$ is an odd number.
We can show that the real parameters $x_i$ are the kink positions 
by substituting $B_0$ into the energy density \eqref{eq:Esosup}
\beq
\mathcal E ~=~ \left\{ \begin{array}{ll} 
\displaystyle \frac{4 \pi m^2}{g^2} \sum_{i=1}^{[n/2]} {\rm sech}^2 \big[ m (x-x_i) \big] & \mbox{for $G=SO(2n)$} \\ 
\displaystyle \frac{2 \pi m^2}{g^2} \sum_{i=1}^{n} {\rm sech}^2 \big[ m (x-x_i) \big] & \mbox{for $G=USp(2n)$} \end{array} \right.. \label{eq:energyUSp}
\eeq
The phases of the monopoles and the orientational moduli of the non-Abelian vortices are contained in the unitary matrix $U$. By assuming that the moduli parameters are dynamical, we can obtain the effective action from the vortex effective action Eq.\,\eqref{eq:USpL}. Here, we restrict ourselves to the case of $USp(4)$ for simplicity. In this case, there exist three types of vortices: one is non-Abelian and the others are Abelian. The kink solution Eq.\,\eqref{eq:USp_kink} represents two monopoles interpolating the three vortices. It is convenient to parameterize the unitary matrix $U \in U(2)$ by two orthogonal unit vectors $\vec n_1$ and $\vec n_2$ in the following way
\beq
U = \Big( \, \vec n_1 \,,\, \vec n_2 \, \Big), \hs{10} |\vec n_1|^2 = |\vec n_2|^2 = 1, \hs{10} \vec n_1^\dagger \cdot \vec n_2 = 0.
\eeq
This $U(2)$ moduli parameters parametrize the phases of the two monopoles and the $\C P^1$ orientation of the non-Abelian vortex. Substituting the kink solution Eq.\,\eqref{eq:USp_kink} into the vortex effective Lagrangian Eq.\,\eqref{eq:USpL}, we obtain the following effective Lagrangian
\beq
\frac{g^2}{4\pi} \mathcal L_{eff} &=& f_1(x) \Big[ m^2 \dot x_1^2 +4 |\vec n_1^\dagger \cdot \p_\alpha \vec n_1 |^2 \Big] + f_2(x) \Big[ m^2 \dot x_2^2 +4 |\vec n_2^\dagger \cdot \p_\alpha \vec n_2 |^2 \Big] \notag \\
&& + f_3(x) \Big| ( \p_\alpha - \vec n_1^\dagger \cdot \p_\alpha \vec n_1) \vec n_1 \Big|^2 + f_4(x) \Big| \vec n_1^\dagger \cdot \p_\alpha \vec n_2 + \vec n_2^\dagger \cdot \p_\alpha \vec n_1 \Big|^2. 
\eeq
Note that $|(\p_\alpha - \vec n_1^\dagger \cdot \p_\alpha \vec n_1) \vec n_1|^2 = |(\p_\alpha - \vec n_2^\dagger \cdot \p_\alpha \vec n_2) \vec n_2|^2$ is the standard $\C P^1$ Lagrangian. The functions $f_i(x)$ and their asymptotic forms 
in the large mass limit $m \rightarrow \infty$ are given by
\beq 
f_1(x) &=& \frac{1}{(e^{m(x-x_1)}+e^{-m(x-x_1)})^2} ~\rightarrow~ \frac{1}{2m} \delta(x-x_1), \\
f_2(x) &=& \frac{1}{(e^{m(x-x_2)}+e^{-m(x-x_2)})^2} ~\rightarrow~ \frac{1}{2m} \delta(x-x_2),
\eeq
\beq
f_3(x) &=& \frac{2 (e^{\frac{m}{2} (x_1-x_2)}-e^{-\frac{m}{2} (x_1-x_2)})^2}{(e^{m(x-x_1)}+e^{-m(x-x_1)})(e^{m(x-x_2)}+e^{-m(x-x_2)})} ~\rightarrow~ 2 \theta(x-x_1) \theta(x_2-x) , \\
f_4(x) &=& \frac{1}{2} \frac{1}{(e^{m(x-x_1)}+e^{-m(x-x_1)})(e^{m(x-x_2)}+e^{-m(x-x_2)})} ~\approx~ \mathcal O( e^{-m|x_1-x_2|} ).
\eeq
Therefore, in the large mass limit $m \rightarrow \infty$, the effective Lagrangian becomes 
\beq
\mathcal L_{eff} &=& \frac{4\pi}{g^2} \Bigg[ \delta(x-x_1) \bigg( \frac{m}{2} \dot x_1^2 + \frac{2}{m} |\vec n_1^\dagger \cdot \p_\alpha \vec n_1|^2 \bigg) + \delta(x-x_2) \bigg( \frac{m}{2} \dot x_2^2 + \frac{2}{m} |\vec n_2^\dagger \cdot \p_\alpha \vec n_2|^2 \bigg) \notag \\
&{}& \hs{8} + 2 \theta(x - x_1)\theta(x_2 - x) \Big| (\p_\alpha - \vec n_1^\dagger \cdot \p_\alpha \vec n_1) \vec n_1 \Big|^2 + \mathcal O(e^{-m|x_1-x_2|}) \Bigg].
\eeq
This effective action is essentially the same as the $U(4)$ case discussed in the previous section (see Eq.\,\eqref{eq:U4}). The difference can be seen when the separation between two monopoles becomes smaller than $m^{-1}$. In the $\C P^{N-1}$ sigma model, kinks have a fixed ordering and their positions cannot be exchanged. For example, the energy density of the kink configuration in the $U(4)$ case is given by
\beq
\mathcal E = \frac{2\pi}{g^2} \p_x^2 \log (e^{m(x-x_1)} + 1 + e^{-m(x-x_2)} ).
\eeq
By fixing the center of mass position $(x_1+x_2)/2$ and taking the limit $x_1-x_2 \rightarrow - \infty$, we can see that the two kinks become a single kink with a larger mass. In this limit, the $S^3$ moduli parameter $\vec n$ disappears from the kink solution Eq.\,\eqref{eq:kink_solU4}, and hence from the effective action Eq.\,\eqref{eq:U4_0}. On the other hand, as we can see from the energy density Eq.\,\eqref{eq:energyUSp}, the positions of the $USp(4)$ kinks can be exchanged. Furthermore, in the the coincident limit $x_1-x_2 \rightarrow 0$, the matrix $X_{USp(2n)}$ in Eq.\,\eqref{eq:X} becomes the identity matrix and hence there are remaining degrees of freedom parameterizing $U(1) \times SU(n)/SO(n) \subset U(n)$. A similar phenomenon has also been seen in the Grassmannian $Gr(2n,n)$ sigma model, where the surviving degrees of freedom on coincident kinks are $U(n) \subset SU(n) \times SU(n) \times U(1)$ \cite{Shifman:2003uh,Eto:2008dm}. This difference between $SU$ and $SO/USp$ cases would play an important role in the discussion of bound states of dyonic vortices and monopoles (kinks) \cite{Eto:2008dm}. 
 
\section{Summary and Discussion}\label{Conclusions}

In this paper, we derived the effective action for the $1/4$ BPS monopole-vortex complex. We studied the interactions between the orientational moduli of non-Abelian vortices and monopoles. We considered $U(N)$, $U(1) \times SO(2n)$ and $U(1) \times USp(2n)$ gauge groups and obtained the effective action for various choices of the hypermultiplet masses. In the limit of large masses, we found that the effective action is expressed by a non-linear sigma model with boundary terms located at monopole positions. Its typical form is:
\beq
S_{eff} &=& \frac{4\pi}{g^2} \int dx dt \, \bigg[ \frac{1}{m} \delta(x-x_1) \left( \frac{m^2}{4} \dot x_1^2 + |\vec n_1^\dagger \cdot \p_\alpha \vec n_1|^2 \right) + \frac{1}{m} \delta(x-x_2) \left( \frac{m^2}{4} \dot x_2^2 + |\vec n_2^\dagger \cdot \p_\alpha \vec n_2|^2 \right) \notag \\
&& \hs{18} + \theta(x-x_2)\theta(x_1-x) | D_\alpha \vec n |^2 \bigg] + \mathcal O \left( \frac{m^2}{g^2 \xi} \right). 
\eeq
where two different parts can be regarded as follows: one is the vortex part, proportional to the product of the two step functions, describing the dynamics of the orientational modes on the vortex worldsheet; the other is the monopole part, proportional to the Dirac delta functions, containing both the kinetic terms of the monopoles and the boundary terms for the orientational moduli. This particular action describes a complex of three vortices, of which one is non-Abelian, as depicted in Fig. \ref{NANApicture}. This action would be useful for the understanding of quantum physics of non-Abelian monopoles with zero modes arising due to non-Abelian symmetry. 

In this paper, we have assumed that the mass scale $m$ is much smaller than $g \sqrt{\xi}$ and our effective action is the leading order approximation with respect to the ratio $m^2/(g^2 \xi)$. In addition to higher derivative corrections, the subleading $\mathcal O(m^2/(g^2 \xi))$ correction to the vortex worldsheet theory has been obtained \cite{Eto:2012qd}. In principle, the method used in Ref.\,\cite{Eto:2012qd} is also applicalbe to the case of the 1/4 BPS monopole-vortex complex discussed in this paper. Furthermore, it has been shown that the subleading correction to the vortex worldsheet theory does not modify the static BPS kink solutions. Thus, it is interesting to see if there exist higher order corrections to the interaction terms between monopoles and orientational moduli.

Although we have determined only the bosonic part of the effective action in this paper, there should also be terms containing fermionic degrees of freedom. Those fermionic zero modes can be determined by solving the equations of motion for the fermionic fields and their effective action can be obtained similarly to the case of the bosonic degrees of freedom. Since the configurations we have discussed are 1/4 BPS states in $\mathcal N=2$ theories, their effective action should have two real supercharges. It is interesting to see how our effective action, which consists of 2-dimensional sigma model and boundary terms, is generalized to a supersymmetric theory. Such a supersymmetric effective theory will be useful in undestanding the role of the non-Abelian zero modes (orientational moduli) in the quantum theory of non-Abelian monopoles.


\end{document}